\documentclass[aps,twocolumn,pra,showpacs]{revtex4}
\usepackage{graphicx}

\newcommand{\be}{\begin{equation}}
\newcommand{\ee}{\end{equation}}
\newcommand{\bea}{\begin{eqnarray}}
\newcommand{\eea}{\end{eqnarray}}

\begin{document}

\title{Towards single-atom detection on a chip}

\author{Peter Horak, Bruce G.\ Klappauf}
\affiliation{Optoelectronics Research Centre, University of
Southampton, Southampton SO17 1BJ, UK}
\author{Albrecht Haase, Ron Folman, J\"org Schmiedmayer}
\affiliation{Physikalisches Institut, Universit\"at Heidelberg,
D-69120 Heidelberg, Germany}
\author{Peter Domokos}
\affiliation{Institut f\"ur Theoretische Physik, Universit\"at
Innsbruck, Technikerstr.\ 25, A-6020 Innsbruck, Austria}
\author{E.\ A.\ Hinds}
\affiliation{Sussex Centre for Optical and Atomic Physics,
University of Sussex, Brighton BN1 9QH, UK}

\date{\today}


\begin{abstract}
We investigate the optical detection of single atoms held in a
microscopic atom trap close to a surface. Laser light is guided by
optical fibers or optical micro-structures via the atom to a
photo-detector. Our results suggest that with present-day
technology, micro-cavities can be built around the atom with
sufficiently high finesse to permit unambiguous detection of a
single atom in the trap with 10 $\mu$s of integration. We compare
resonant and non-resonant detection schemes and we discuss the
requirements for detecting an atom without causing it to undergo
spontaneous emission.
\end{abstract}

\maketitle

\section{Introduction}

The subject of matter wave optics is advancing rapidly, driven
both by the fundamental interest in quantum systems and by the
prospect of new instruments based on the quantum manipulation of
neutral atoms. The recent miniaturization of atom traps above
microfabricated surfaces \cite{hindsreview,chipreview} has opened
the possibility of using neutral atoms to perform quantum
information processing (QIP) on a chip. This technology is
attractive because it appears robust and scalable, and because
trapped neutral atoms can have long coherence times.

Experiments have shown that it is possible to trap, guide, and
manipulate cold, neutral atoms in miniaturized magnetic traps
above a substrate using either microscopic patterns of permanent
magnetization in a film or microfabricated wire structures
carrying current or charge. These atom chips can create potentials
where atoms are confined strongly enough  to consider implementing
quantum logic gate schemes \cite{tommaso}. In the last year,
several groups have been able to load atom chip microtraps with
Bose-Einstein condensates (BEC) \cite{BECchip}, which may serve as
a coherent source of qubits. A next important step for QIP is the
detection of individual atoms on a chip with a signal-to-noise
ratio better than unity in, say, 10 $\mu$s. Here we propose that
this can be achieved using very small optical cavities
microfabricated on the chip, in which the presence of a single
atom produces a sufficient response in the light field  to permit
detection.

Section \ref{sec:model} presents a generic mathematical model of
atom detection in a cavity. In Secs.\ \ref{sec:detector} and
\ref{sec:homodyne} we discuss resonant and off-resonant detection.
Section \ref{sec:motion} deals with the forces acting on the atom
due to the detection process.  We give a numerical simulation of
an atom crossing the detection cavity while moving in an atom
guide. In Sec.\ \ref{sec:designs} we discuss some practical
aspects of making optical cavities and waveguides on an atom chip.
We conclude in Sec.\ \ref{sec:conclusions}. Appendix
\ref{sec:modes} provides details of the particular model cavity
used in our analysis.

%%%%%%%%%%%%%%%%%%%%%%%%%%%%%%%%%%%%%%%%

\section{Generic model}
\label{sec:model}

In this section we develop a model for a two-level atom coupled to
the coherent light field in a microscopic cavity. The picture is
essentially one already used to describe  experiments in the
strong coupling regime of cavity quantum electrodynamics (QED)
\cite{Berman,Kimble,Rempe}, however, we are interested here in a
different region of parameter space. Because the cavities of
interest are small, with mirrors of limited reflectivity, the
cavity decay rates $\kappa$ are many orders of magnitudes faster
than those of the best optical resonators. On the other hand,
these short cavities can support stable modes that have extremely
small waist size ($\sim $ 1 $\mu$m), resulting in very strong
atom-cavity coupling $g$. In such cavities even a small photon
number can be sufficient to saturate the atomic transition, so we
need to take nonlinear effects into account. The elements of the
atomic density operator $\rho$ satisfy the optical Bloch equations
 \bea
 \frac{d}{dt}\rho_{01} & = & (-\Gamma-i\Delta_a)\rho_{01}
   +g\alpha^*(\rho_{00}-\rho_{11}) ,\\
 \frac{d}{dt}\rho_{11} & = & -2\Gamma\rho_{11}
   +g(\alpha^*\rho_{10}+\alpha\rho_{01})  .
 \eea
Here, $2\Gamma$ is the decay rate of the excited atomic state,
$\Delta_a$ is the detuning of the driving laser from the atomic
resonance, and $g$ is the single-photon Rabi frequency at the
position of the atom. For simplicity we will assume in the
following that the atom sits at an intensity maximum of the light.
The light field in the cavity is treated classically, i.e., a
coherent state is assumed at all times. The coherent state
amplitude $\alpha$ obeys the equation of motion
 \be
 \frac{d}{dt}\alpha = (i\Delta_c-\kappa)\alpha+g\rho_{10}+\eta.
 \ee
Here $\Delta_c$ denotes the detuning of the driving laser from the
cavity resonance. The decay rate of the cavity field is
$\kappa\equiv\kappa_T+\kappa_{loss}$, made up of $\kappa_T$ due to
photons that pass through the cavity mirrors, and $\kappa_{loss}$
due to photons that leave the cavity by other processes. The term
$\eta$ is the cavity pumping rate, related to the pumping laser
power by $\eta=\sqrt{j_{in}\kappa_T}$, where $j_{in}$ is the rate
of photons incident on the cavity.

The stationary solutions for the light amplitude in the cavity and
for the population of the atomic excited state are
 \bea
 \alpha & = & \frac{\eta}{(\kappa+\gamma)-i(\Delta_c-U)} \label{eq:alphastat},\\
 \rho_{11} & = & \frac{g^2N}{\Delta_a^2+\Gamma^2+2g^2N},
 \eea
where
 \bea
  \gamma & = & \frac{g^2\Gamma}{\Delta_a^2+\Gamma^2+2g^2N} \label{eq:gamma},\\
  U & = & \frac{g^2\Delta_a}{\Delta_a^2+\Gamma^2+2g^2N}\label{eq:U},
 \eea
and $N=|\alpha|^2$ is the mean intracavity photon number. When the
quantity $2 g^2N$ is small (large) compared with $\Gamma^2$, we
say that the atomic saturation is low (high). Note that Eq.\
(\ref{eq:alphastat}) defines the stationary field amplitude only
implicitly because $\gamma$ and $U$ depend on $N$. Hence this
equation normally has to be solved numerically.

The presence of the atom is detected through its effect on the
field amplitude $\alpha$.   This is partly due to the spontaneous
scattering, which adds $\gamma$ to the cavity damping rate in Eq.\
(\ref{eq:alphastat}), and partly to the coherent scattering, which
adds $-U$ to the cavity detuning. With the atom at resonance and
unsaturated, the additional damping is $g^2/\Gamma$ and the ratio
of this to the intrinsic cavity damping is $g^2/(\Gamma\kappa)$,
the cooperativity parameter of laser theory. This parameter is
fundamental to the description of the atom-cavity interaction.
When it is much smaller (larger) than unity, we describe the
atom-cavity coupling as weak (strong). For a cavity of length $L$,
it can be expressed as
 \be
 \frac{g^2}{\Gamma\kappa} = 2\frac{\sigma_a}{A} n_{rt}
 \label{eq:g2kappa}
 \ee
where $\sigma_a=3\lambda^2/(2\pi)$ is the resonant atomic
interaction cross section for light of wavelength $\lambda$, $A$
is the cross section of the cavity mode at the position of the
atom, and $n_{rt}=c/(4L\kappa)$ is the average number of round
trips of a cavity photon before its decay. (Provided the
reflectivity of the mirrors is close to 1, the finesse of the
cavity is just $4 \pi n_{rt}$). This agrees with the naive
expectation that the effect of the atom should depend on the
fraction of the light within its cross section and should increase
linearly with the number of times each photon passes the atom.
Note however that Eq.\ (\ref{eq:g2kappa}) is restricted to beam
waist sizes $A$ that are more than a few times the cross section
$\sigma_a$ because it only holds within the dipole and paraxial
approximations \cite{vanEnk,domokos02}.  Let us emphasize some
important scaling properties of this quantity in the following.

(i) If the cavity decay rate is dominated by losses at a fixed
number of material interfaces, then the number of round trips
$n_{rt}$ is independent of the cavity length.  It follows from
Eq.\ (\ref{eq:g2kappa}) that if the atom is at the center of the
cavity and the waist area $A$ is held constant, then
$g^2/(\Gamma\kappa)$ is also independent of the cavity length. In
this case, if one scales the cavity length and detuning by
$L\rightarrow rL$ and $\Delta_c\rightarrow\Delta_c/r$ but keeps
the pump laser power and $\Delta_a$ constant, then the detector
signal-to-noise ratio (as discussed in the following sections) and
the back-action on the atom during the measurement will remain
unchanged.

(ii) Equation (\ref{eq:g2kappa}) suggests that, within the
paraxial approximation, the beam cross section at the atomic
position should be as small as possible to increase the coupling
of the atom to the light. However, the beam divergence increases
as the waist is made smaller, increasing the diffractive losses
and other non-paraxial imperfections and reducing $n_{rt}$.
Consequently the optimum value for $A$ depends on the specific
details of the cavity and its losses.

In the following two sections we will discuss two possible ways to
detect the presence of an atom by measuring the output light beam.
The first is to measure a dip in the output intensity using pump
light that is resonant with the atom.  The second is to measure
the phase shift of the output light using an off-resonant pump.
The more general case, in which an atom changes both the amplitude
\textit{and} the phase of the output beam, can be qualitatively
understood by considering these two extremes.

%%%%%%%%%%%%%%%%%%%%%%%%%%%%%%%%%%%%%%%%

\section{Resonant atom detection}
\label{sec:detector}

Let us compare the number of resonant photons transmitted in a
time $\tau$ through the output mirror to a detector, with and
without an atom in the cavity. For a given intra-cavity photon
number $N$ in a symmetric cavity, the number of photons arriving
at the detector is $N_{out}=N\kappa_T\tau$. (This can be enhanced
to $2N\kappa_T\tau$ if the input mirror has much higher
reflectivity than the output mirror). We are thus interested in
the difference $N_{out,0}-N_{out}$ where $N_{out,0}$ is the output
of the empty cavity. This signal must be compared to the quantum
noise of the measurement, i.e.\ to the width $\sqrt{N_{out}}$ of
the Poissonian photon number distribution of a coherent state. The
signal-to-noise ratio of this measurement scheme is therefore
 \be
 S = \frac{N_{out,0}-N_{out}}{\sqrt{N_{out}}}.
 \label{eq:Sres}
 \ee
We will be interested in the regime $N_{out,0}>1$. With the cavity
and the atom both at resonance, the atomic transition saturates at
very low intra-cavity photon number making the difference signal
weak, so it is natural to consider using off-resonant excitation
at higher intensity. However, it can be shown that $S$ is maximum
for resonant pumping and we will therefore pursue the idea of
using $\Delta_c=\Delta_a=0$ in the following. Sensitive detection
with detuning requires a more sophisticated homodyne technique,
which we discuss in Sec.~\ref{sec:homodyne}.

With $\Delta_c=\Delta_a=0$, analytical solutions of Eq.\
(\ref{eq:Sres}) can be found for some limiting cases. In the limit
of low atomic saturation, where $2g^2N\ll\Gamma^2$, we find

\be
 S = \sqrt{j_{in}\tau}\frac{g^2}{\kappa\Gamma}
     \frac{\kappa_T}{\kappa}\times
 \left\{
 \begin{array}{ll}
 2  \\
 1
 \end{array}
\mbox{ for }\left(\frac{g^2}{\kappa\Gamma}\right)
\begin{array}{ll}
 \ll 1   \\
 \gg 1
 \end{array}
 \right.
 \label{eq:Sweak}
 \ee
Hence the signal-to-noise ratio at low saturation increases with
the square root of the incident laser power and integration time,
but linearly with the number of photon round trips in the cavity.
Losses also degrade the sensitivity through the factor
$\kappa_T/\kappa$. In the opposite limit of strong saturation,
where $2g^2N\gg\Gamma^2$, Eq.\ (\ref{eq:alphastat}) yields
 \be
 N=\frac{\eta^2}{\kappa^2}-\frac{\Gamma}{\kappa}
 \label{eq:Nstrong}
 \ee
which gives
 \be
 S = \Gamma\sqrt{\frac{\tau}{j_{in}}}.
 \label{eq:Sstrong}
 \ee
Thus the signal-to-noise ratio \textit{decreases} with the square
root of the intensity when the transition is saturated. This is
because the noise due to fluctuations of $N_{out}$ approaches
$\sqrt{j_{in}\tau}$, whereas the number of photons scattered by
the atom is limited to $\Gamma\tau$. This result is independent of
all the cavity parameters.

\begin{figure}[tb]
 \includegraphics[height=4.5cm]{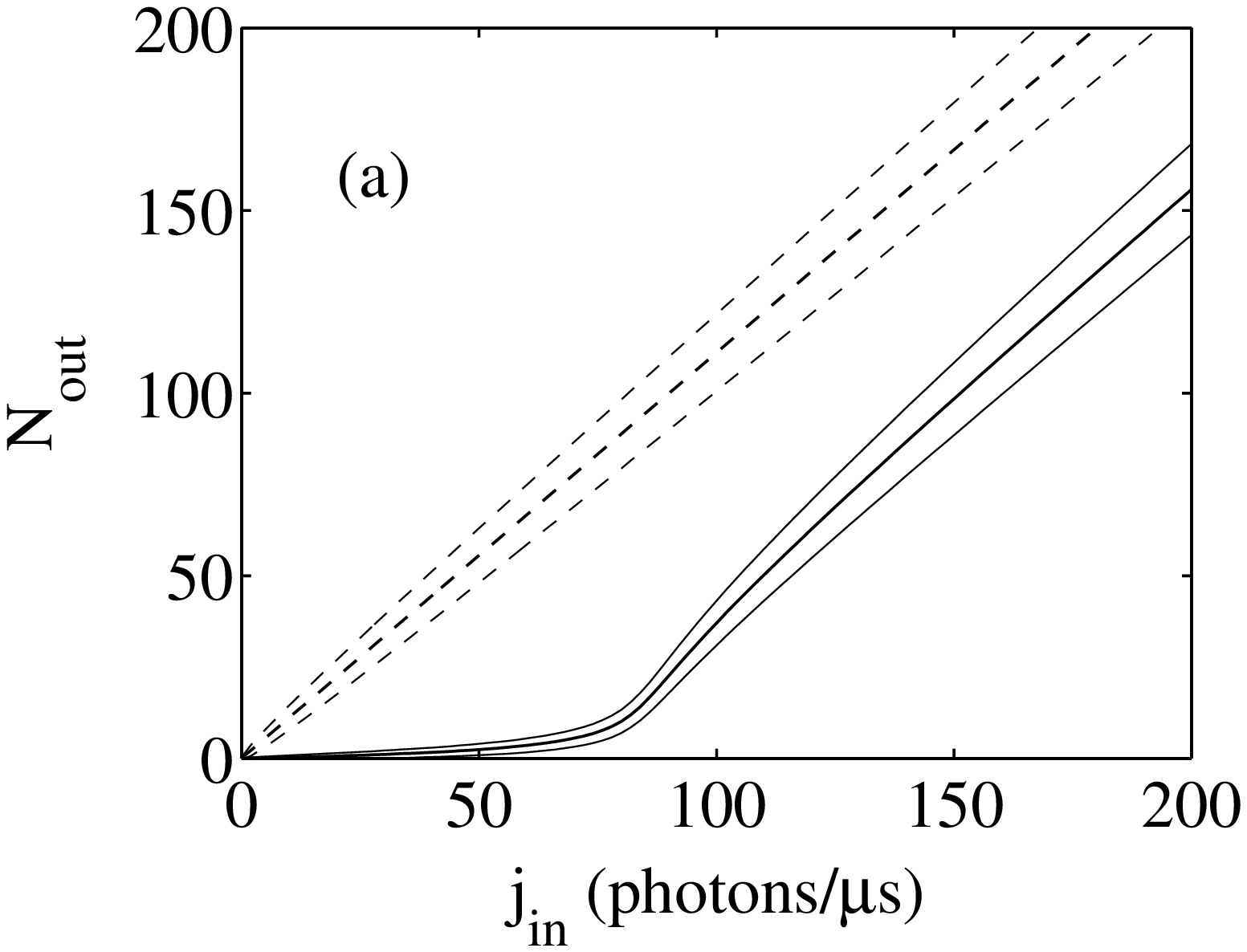}
 \includegraphics[height=4.5cm]{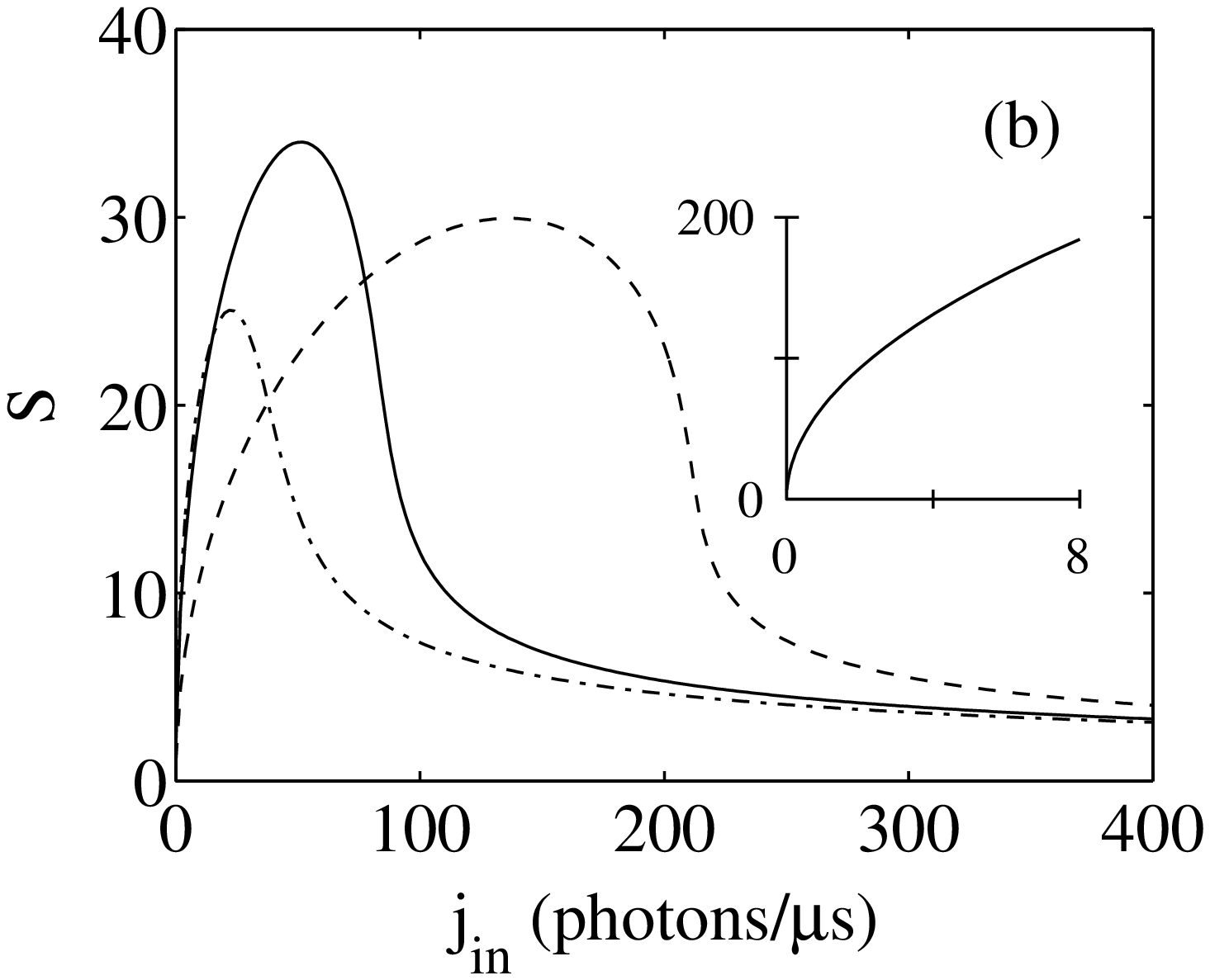}
 \includegraphics[height=4.5cm]{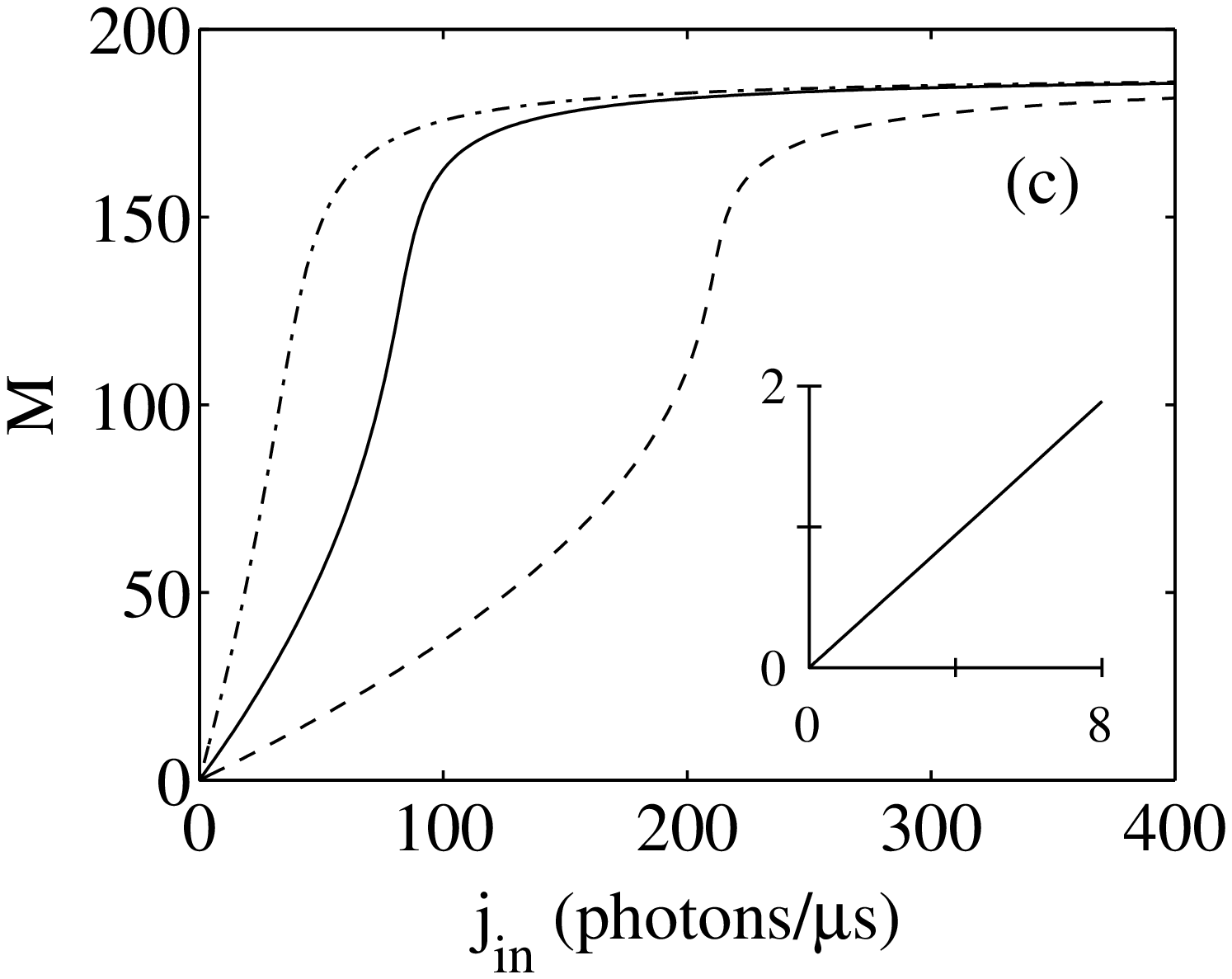}
 \caption{Resonant detection of a single Rb atom with
 $(g,\Gamma,\kappa_{loss})=2\pi\times(12,3,6)$ MHz, $\tau=10\,\mu$s.
 (a) Number $N_{out}$ of photons transmitted through the cavity
 with (solid line) and without (dashed) atom. Thin lines
 correspond to shot noise, $\kappa_T=2\pi\times 3$ MHz.
 (b) Signal-to-noise ratio $S$ for $\kappa_T/(2\pi)=1$ MHz
 (dashed), 3 MHz (solid), and 10 MHz (dash-dotted). Inset: $S$ for a
 better cavity with $\kappa_{loss}=\kappa_T=2\pi\times 0.59$ MHz.
 (c) Number $M$ of photons emitted spontaneously by the atom during the measurement
 [parameters as in (b)]. The parameters are justified by calculations
 performed in appendix \ref{sec:modes} for an experimentally feasible cavity.
 \label{fig:noloss}}
\end{figure}

In Figures \ref{fig:noloss} and \ref{fig:loss} we show some
numerical results based on resonant pumping of the simple cavity
described in Appendix \ref{sec:modes}. Figure \ref{fig:noloss}(a)
shows the output photon number as a function of the pump power,
both with and without an atom in the cavity. With an empty cavity,
the output is proportional to the pump power for all parameters.
With an atom in the cavity this scaling also holds as long as the
atomic saturation is small. For the parameters chosen here the
atom-light coupling is strong, i.e., $g^2 \gg \kappa\Gamma$, and
thus the atom significantly reduces the intra-cavity photon number
as long as it is not saturated. For strong atomic saturation we
find that $N_{out,0}$ and $N_{out}$ differ by a constant value, in
accordance with the second term of Eq.\ (\ref{eq:Nstrong}). In
Fig.\ \ref{fig:noloss}(b) we plot the signal-to-noise ratio $S$
versus pumping for three different values of $\kappa_T$. In each
case, $S$ increases with $\sqrt{j_{in}}$ for weak fields and then
decreases in stronger fields in accordance with Eqs.\
(\ref{eq:Sweak}) and (\ref{eq:Sstrong}). The optimum sensitivity,
observed close to atomic saturation  ($2Ng^2 \approx \Gamma^2)$,
is obtained at a different pump power on each curve. There is also
an optimum value for the mirror transmission. We found numerically
that this occurs at $\kappa_T\approx\kappa_{loss}/2$ with strong
coupling and $\kappa_T\approx\kappa_{loss}$ with weak coupling.
For the parameters given, a single atom can be detected with
signal-to-noise ratios of up to 35.

Figure \ref{fig:noloss}(c) shows $M \equiv 2\Gamma\tau\rho_{11}$,
the number of photons spontaneously scattered by the atom during
the detection process. With the parameters used here, smaller
$\kappa_T$ gives smaller $M$ at a given pump power, but this
depends on the value of $\kappa_{loss}$ and is not always so. For
example, $M\propto 1/\kappa_T$ when $\kappa_{loss}=0$ and the
coupling is weak. In the strongly saturated regime, all curves
converge to the limit $M=\Gamma\tau$. This spontaneous scattering
causes momentum diffusion of the atom and loss of atomic
coherences.  For the purpose of atom detection, this does not
constitute a problem.  However, if one had in mind to use the
atom-cavity coupling for reversible, quantum logic operations it
would be essential to have little or no spontaneous decay. One
would then approach the regime of so-called interaction-free
measurements \cite{Elitzur,Kwiat,Karlsson}. Atomic motion will be
discussed in more detail in Sec.\ \ref{sec:motion}. In the weak
saturation regime we can use Eq.\ (\ref{eq:Sweak}) to express $M$
as a function of the signal-to-noise ratio, \be
 M = S^2\frac{\kappa}{\kappa_T}\times
 \left\{
 \begin{array}{ll}
 \frac{1}{2}\left(\frac{g^2}{\kappa\Gamma}\right)^{-1}  \\
 2\left(\frac{g^2}{\kappa\Gamma}\right)^{-3}
 \end{array}
\mbox{ for }\left(\frac{g^2}{\kappa\Gamma}\right)
\begin{array}{ll}
 \ll 1   \\
 \gg 1
 \end{array}
 \right.
 \label{eq:Mweak}
 \ee
Hence with weak coupling and fixed signal-to-noise ratio, the
number of spontaneously scattered photons is inversely
proportional to the number of round trips $n_{rt}$. This is
because the probability of the atom scattering a given photon is
directly proportional to $n_{rt}$. In the strong coupling regime,
on the other hand, the behavior is nonlinear and $M\propto
1/n_{rt}^3$. Therefore an increase of $n_{rt}$ by a factor of 10
would reduce the photon scattering by three orders of magnitude.
Of course, the increase of $n_{rt}$ also reduces the number of
photons in the cavity output. When $\kappa_{loss}$ and $\kappa_T$
are reduced to $2\pi\times 0.59$ MHz and $j_{in}=2$
photons/$\mu$s, we find $N_{out,0}=5$ and $M=0.47, S=93$, as shown
in the insets of Figs.\ \ref{fig:noloss}(b) and (c).

\begin{figure}[tb]
 \includegraphics[height=4.5cm]{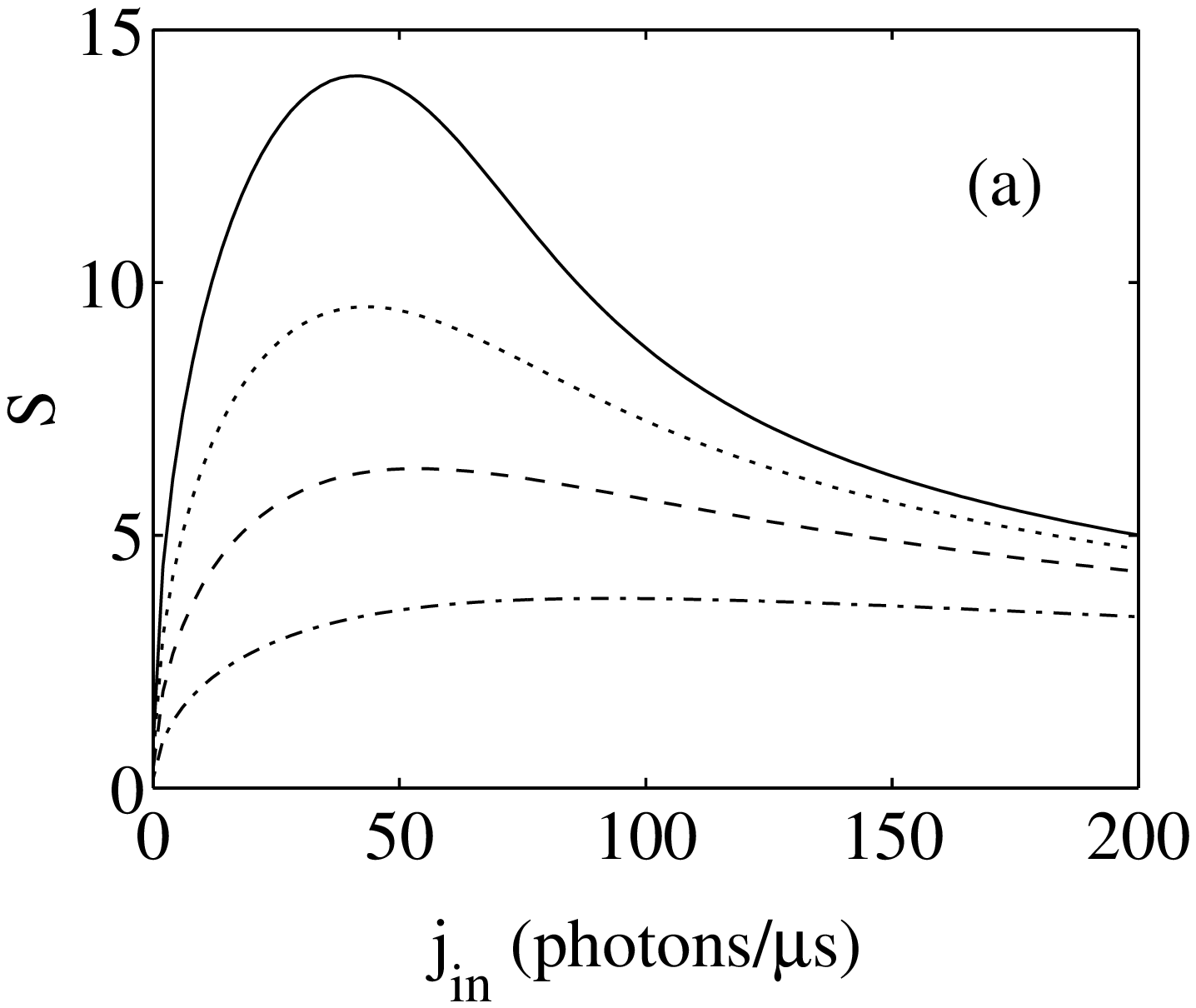}
 \includegraphics[height=4.5cm]{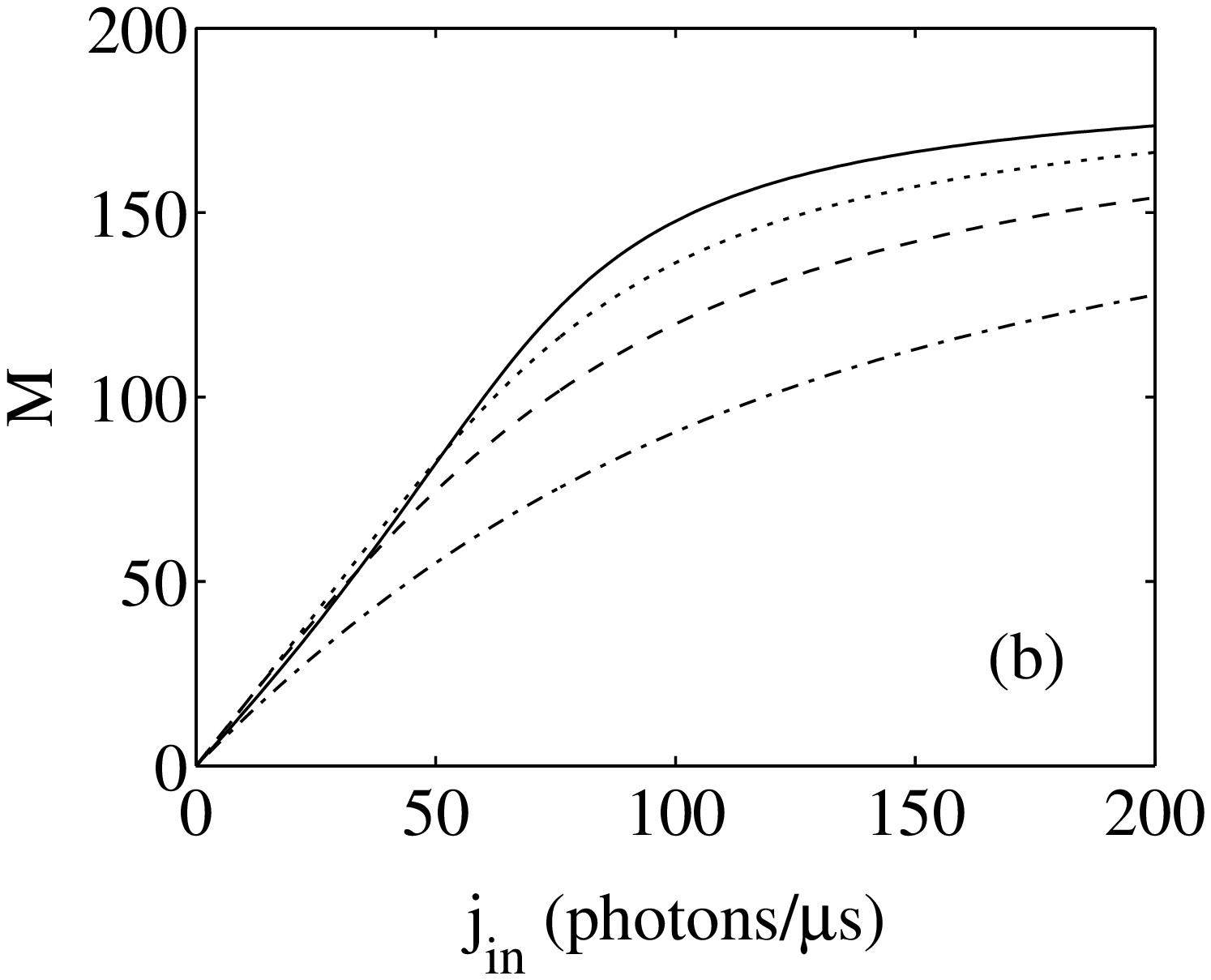}
 \caption{Same as Fig.~\ref{fig:noloss} but with increased loss
 rates $\kappa_{loss}/(2\pi)=14$ MHz (solid line), 22 MHz
 (dotted), 38 MHz (dashed), 86 MHz (dash-dotted). Again
 $\tau=10\,\mu$s. For each curve $\kappa_T=\kappa_{loss}/2$ to
 get optimum signal-to-noise ratios.
 \label{fig:loss}}
\end{figure}

Figure \ref{fig:loss} is similar to Fig.~\ref{fig:noloss} but with
four larger values of $\kappa_{loss}$ and with $\kappa_T$ set
equal to $\kappa_{loss}/2$ in each case so as to achieve optimum
signal-to-noise ratio. The curves for $S$ and $M$ as a function of
the pump power are generally similar to Fig.\ \ref{fig:noloss},
but in accordance with the lower cavity finesse, the maximum
values of $S$ are reduced. These curves correspond approximately
to additional losses of 1\%, 2\%, 4\%, and 10\% per round trip of
the cavity, as described in Appendix \ref{sec:modes}. For the
largest loss rate, $\kappa_{loss}=2\pi\times 86$ MHz  and the
maximum value of $S$ is 3.75. At this point the number of
scattered photons is $M=86$, which amounts to a considerable
disturbance of the atom. For lower loss rates, on the other hand,
large signal-to-noise ratios can be achieved with only few photon
scattering events.

The possibility of a large signal-to-noise ratio with few
scattered photons makes this cavity-based detection method more
attractive than standard resonance fluorescence imaging. Consider
a simple resonance fluorescence setup where two light guides are
mounted on the chip at $90$ degrees to each other. One guides
laser light to the atom trapped above the chip, while the second
receives scattered photons and conveys them to a photodetector.
The second guide collects only a small fraction of the scattered
light since its end subtends a small solid angle at the atom. For
example, a guide with a 10\ $\mu$m core mounted 10\ $\mu$m away
from the atom collects $\sim$ 5\% of the total scattered photons.
Hence an atom scatters at least 20 photons for every signal photon
and there is no possibility of measuring the atom without
disturbing it strongly.

%%%%%%%%%%%%%%%%%%%%%%%%%%%%%%%%%%%%%%%%%%%%%%%%%%%

\section{Off-resonant detection: homodyne measurement}
\label{sec:homodyne}

In the previous section the cavity and the atom were pumped
resonantly.  We found in that case that atom detection without
spontaneous scattering involves a very small cavity output except
when the coupling is very strong. One might therefore suspect that
the detection scheme could be improved by working with far
off-resonant light, using the dispersive interaction to produce an
optical phase shift. In this section we investigate that idea. We
continue to take $\Delta_c=0$, but now assume a large atom-pump
detuning $\Delta_a\gg\Gamma$ so that the scattering rate $\gamma$
is much less than the light shift $U$ [Eqs.\
(\ref{eq:gamma}),(\ref{eq:U})]. In this situation the dominant
effect of the atom is to shift the resonance frequency of the
cavity, thereby changing the phase of the cavity output but not
its amplitude. Equation (\ref{eq:alphastat}) can then be written
as
 \be
 \alpha \approx \frac{\eta}{\kappa+i U}
        = \frac{\eta}{\kappa}\frac{1}{1-i\phi}
        \approx \frac{\eta}{\kappa}e^{i\phi}
 \ee
where the phase shift
 \be
 \phi = -\frac{U}{\kappa}
 \ee
is assumed to be $\ll 1$.

A balanced homodyne detection scheme can measure the phase of the
cavity output. The cavity output field is mixed with a strong
local oscillator laser field on a 50-50 beam-splitter and the
difference of the light intensities in the two beam-splitter
output ports is measured. The quantum noise of the signal is
determined by the noise of the strong local oscillator, the
signal-to-noise ratio being given by
 \be
 S_{hom} = 2\sqrt{N_{out}}|\sin\phi|
    \approx 2\sqrt{N_{out}}\frac{|U|}{\kappa}.
 \ee
Because of the condition $|\phi|\ll 1$,
$N_{out}=N_{out,0}=j_{in}\tau(\kappa_T/\kappa)^2$. In the limit of
low atomic saturation, we find
 \be
 S_{hom} =
 2\sqrt{j_{in}\tau}\frac{\kappa_T}{\kappa}\frac{g^2}{\Delta_a\kappa}.
 \ee
Hence according to Eq.\ (\ref{eq:g2kappa}), the signal-to-noise
ratio increases linearly with $n_{rt}$. Note also that $S_{hom}$
is $\Gamma/\Delta_a$ times the $S$ of Eq.\ (\ref{eq:Sweak}) for
resonant detection; with the same pump strength the off-resonant
$S_{hom}$ is much smaller than the resonant $S$. For strong atomic
saturation on the other hand, we obtain
 \be
 S_{hom} = \Delta_a\sqrt{\frac{\tau}{j_{in}}}
 \ee
which is larger than the corresponding result for resonant
detection, Eq.\ (\ref{eq:Sstrong}), by a factor of
$\Delta_a/\Gamma$. Again, this is independent of $\kappa$,
$\kappa_T$, $A$, and $n_{rt}$.

The number of photons scattered spontaneously by the atom during
the interaction time can be expressed in terms of $S_{hom}$ as
 \be
 M = S_{hom}^2\frac{\kappa}{\kappa_T}\frac{1}{2}
    \left(\frac{g^2}{\Gamma\kappa}\right)^{-1}.
 \label{eq:Mhom}
 \ee
Therefore, in order to achieve a certain signal-to-noise ratio,
$M$ is the same for the homodyne detection scheme as it is in the
weak-coupling limit of resonant detection (see Eq.\
\ref{eq:Mweak}). Note that the condition $|\phi|\ll 1$ prevents us
from reaching the nonlinear regime of strong coupling here. The
number of photons transmitted through the cavity during this atom
detection is larger by a factor of $\Delta_a^2/\Gamma^2$ than it
is in the resonant detection scheme:
 \be
 N_{out} = \frac{1}{4}S_{hom}^2\left(\frac{\Gamma\kappa}{g^2}\right)^2
           \frac{\Delta_a^2}{\Gamma^2}.
 \ee

\begin{figure}[tb]
 \includegraphics[height=4.5cm]{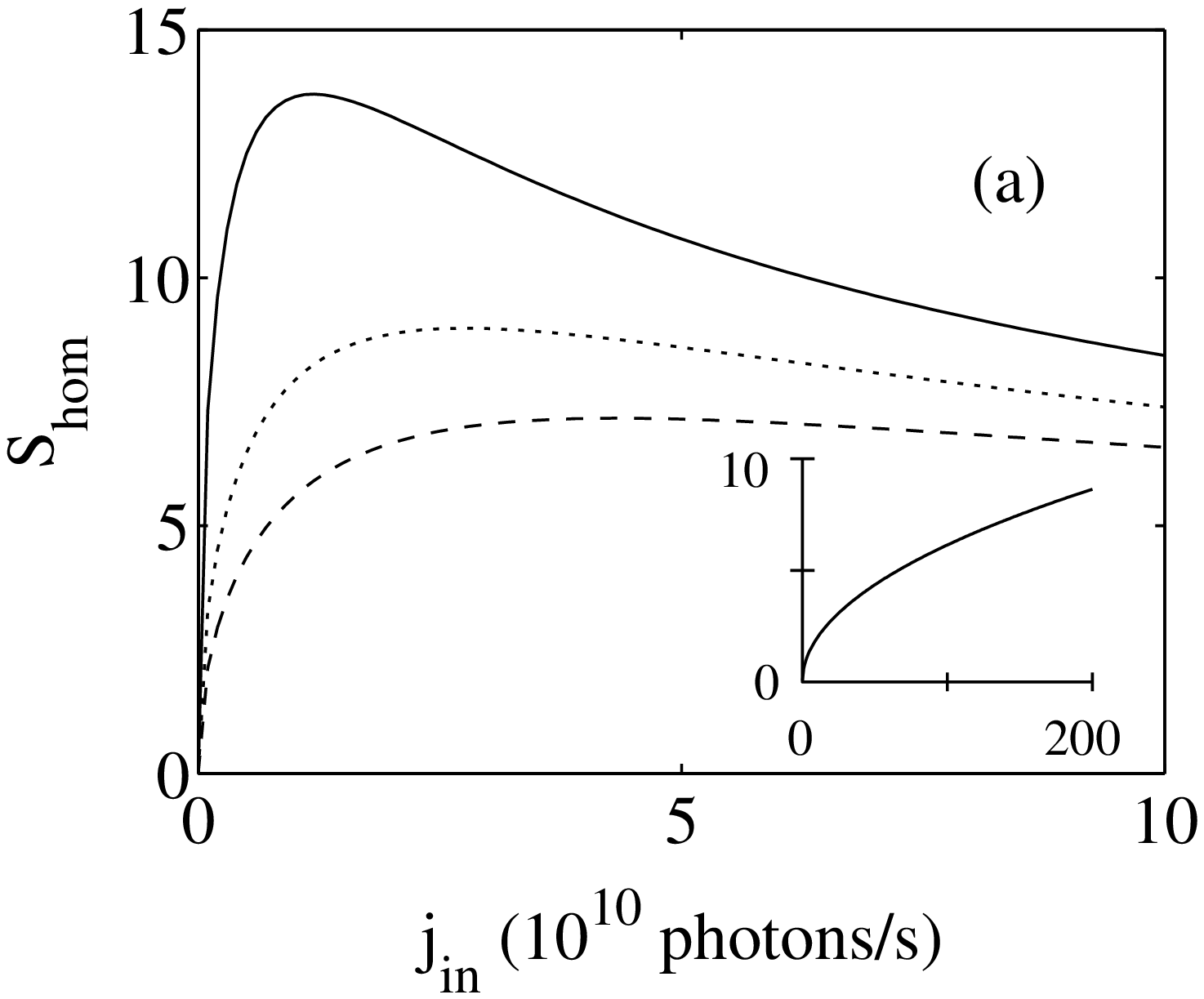}
 \includegraphics[height=4.5cm]{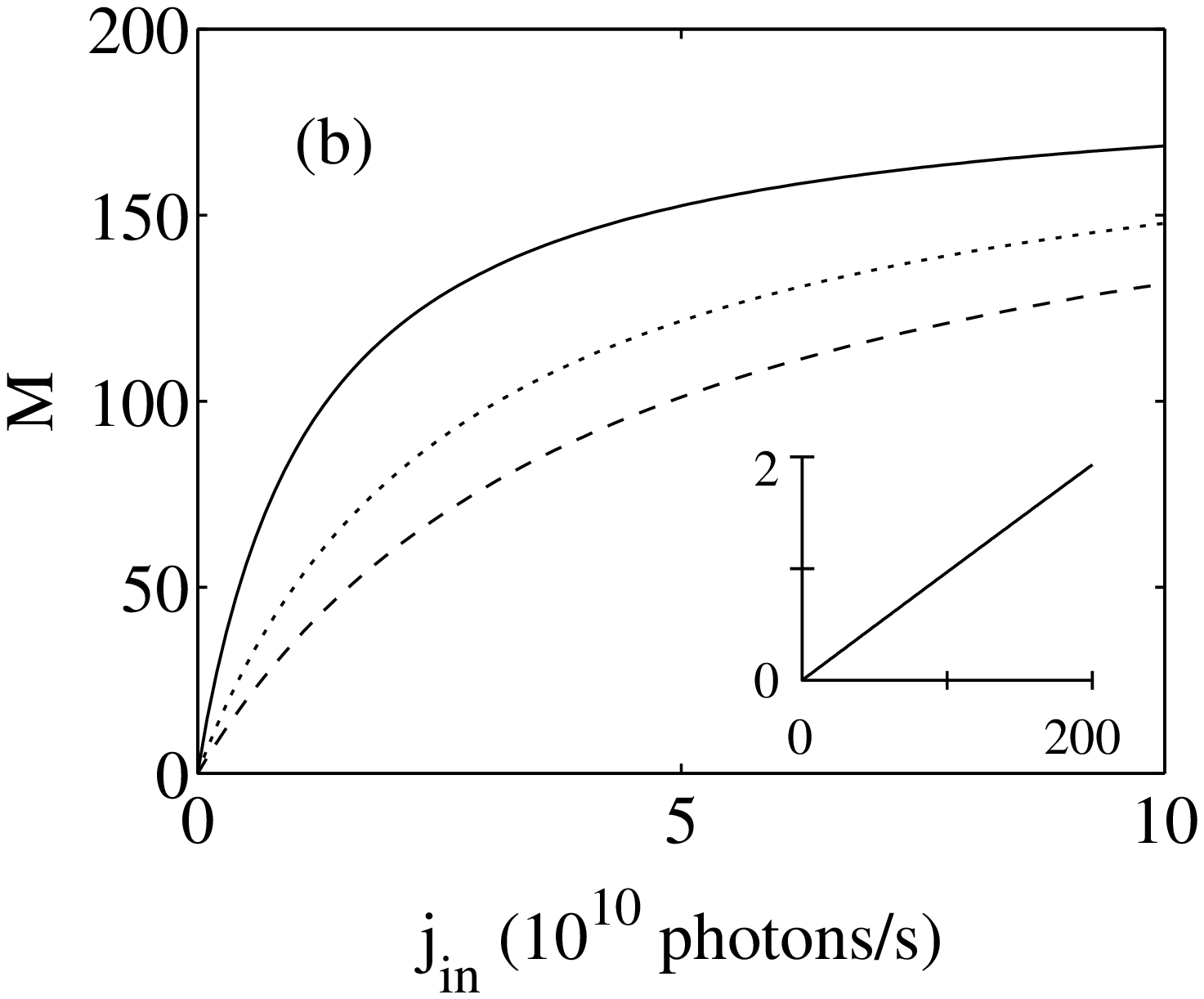}
 \caption{Off-resonant atom detection using a homodyne
 measurement over $\tau=10\,\mu$s. (a) Signal-to-noise ratio
 $S_{hom}$ and (b) number $M$ of spontaneously scattered photons for
 $\kappa_{loss}/(2\pi)=6$ MHz (solid line), 14 MHz (dotted), and 22
 MHz (dashed). For each curve $\kappa_T=\kappa_{loss}$.
 $\Delta_a=50\Gamma$, other parameters as in Fig.\
 \ref{fig:noloss}. Insets:
 $\kappa_{loss}=\kappa_T=2\pi\times0.59$ MHz, $\Delta_a=200\Gamma$.
 \label{fig:hom}}
\end{figure}

Figure \ref{fig:hom}(a) shows the homodyne detection
signal-to-noise ratio $S_{hom}$ as a function of the pump power.
Comparing this with Fig.\ \ref{fig:noloss}(b), we see that the
maximum signal-to-noise ratio for the homodyne detection is
smaller but of the same order of magnitude as for the resonant
detection scheme.  This agrees with the discussion above. The
reduction by about a factor of two for the solid curve (lowest
cavity loss rate) is due to the fact that we are limited to the
weak coupling regime here and therefore do not benefit from the
nonlinear effects of the strong coupling seen in Fig.\
\ref{fig:noloss}(b). Moreover, we find numerically that $S_{hom}$
is maximized for $\kappa_T\approx\kappa_{loss}$, i.e., for
slightly larger mirror transmissions than for the resonant scheme.
However, the pump intensity corresponding to maximum $S_{hom}$ is
much larger because of the large atom-pump detuning $\Delta_a$.
Indeed, the main advantage of the homodyne detection scheme is
that the signal consists of arbitrarily many photons compared to
the few photons in the cavity output for some parameter regimes of
Figs.~\ref{fig:noloss} and \ref{fig:loss}. In Fig.\
\ref{fig:hom}(b) we depict the corresponding number $M$ of
spontaneously scattered photons. Again, $M$ is found to be of the
same order of magnitude as for the resonant detection scheme. This
is because the reduced atom-photon coupling in the far-detuned
limit is compensated by the larger number of photons used to
achieve a good signal-to-noise ratio. The heating of atomic motion
and loss of atom coherence due to photon scattering are therefore
similar in both the resonant and the off-resonant detection
schemes. It follows from Eq.\ (\ref{eq:Mhom}) that a large ratio
of $g^2/(\kappa\Gamma)$ is needed to make $S_{hom}>1$ and $M<1$.
With $\kappa_{loss}=\kappa_T=2\pi\times 0.59$ MHz and
$\Delta_a=200\Gamma$, for example, a pump current of 50
photons/$\mu$s gives $N_{out,0}=125$ and $M=0.49$, $S_{hom}=4.3$
as shown in the insets of Fig.\ \ref{fig:hom}.

%%%%%%%%%%%%%%%%%%%%%%%%%%%%%%%%%%%%%%%%%%%%%%%%%%%

\section{Atomic motion}
\label{sec:motion}

Up to this point we have supposed that the atom is held by the
magnetic microtrap at a maximum of the cavity field for the
duration of the measurement time $\tau$. In this section we show
that reliable detection is possible, even when the atom is allowed
to move around within the light field. We consider a simple
experiment in which the pump laser is continuously on, while
single atoms traverse the cavity at random and have to be detected
during their limited interaction time with the cavity field.
Rubidium atoms trapped in an atom chip waveguide at a temperature
of $1\mu K$ move at typical thermal velocities of 1cm/s.  Such an
atom would typically take $300\mu$s to cross a $3\mu$m cavity mode
waist if it were not interacting with the light.  This is plenty
of time to allow detection, being much longer than the $\tau =
10\mu$s interaction time assumed in previous sections of this
paper. However, the interaction may impart some random momentum
kicks to the atom. If this heating is too great, the atom may
leave the interaction region before it can be detected, regardless
of its initial velocity.

The broadening of the momentum distribution $\Delta p$ over a time
$\tau$ is related to the momentum diffusion coefficient $D$
through the relation $(\Delta p)^2=2D\tau$. For an atom in a
standing wave cavity in the limit of weak atomic saturation
($2g^2N\ll\Gamma^2$) we can use the value of $D$ given in Refs.\
\cite{cavcool}. For simplicity we restrict the following
discussions to the resonant case $\Delta_a=\Delta_c=0$. Thus
 \be
 D = \Gamma(\hbar k)^2 \frac{\eta^2 g^2}{\left[
       \Gamma\kappa + g^2 \cos^2(kz)
     \right]^2},
 \label{eq:D}
 \ee
where $k$ is the wave vector of the light field, $g$ is again the
maximum Rabi frequency, and $z=0$ is at an antinode of the field.
When $z=0$, $2 D \tau$ is equal to $2\Gamma(\hbar k)^2
\rho_{11}\tau$, which is just $(\hbar k)^2 M$. Hence, the momentum
distribution of an atom at an antinode broadens by $\Delta p=\hbar
k \sqrt{M}$ during the interaction time. At this position, the
condition $M<1$ for reversible atom-light interaction is also the
condition for leaving the motional state of the atom unchanged.
Away from the antinodes, however, this is no longer true. For
example, $M=0$ at a node, while $D$ is maximum. In this case, it
is \textit{coherent} scattering of photons between the forward and
backward directions in the cavity that gives rise to the momentum
diffusion \cite{Gordon}. However, in general the use of a cavity
in atom detection reduces the heating of the atom significantly
compared with simple resonance fluorescence detection. Moreover,
the counter-propagating fields in a cavity produce a much smaller
mean scattering force than a single travelling beam, whose
radiation pressure force can expel the atom quickly from the
interaction region. In the ultimate limit of very large
atom-cavity coupling (``interaction-free'' measurement), the
atomic motional state is completely unperturbed by the detection.

Let us now take this position dependence into account by simply
averaging the results of Sec.\ \ref{sec:detector} over the
position along the cavity axis assuming a flat spatial
distribution of the atom. For the spatially averaged
signal-to-noise ratio $\bar{S}$ at resonance we find
 \bea
 \bar{S} &=&  \sqrt{j_{in}\tau}\frac{\kappa_T}{\kappa}
    \left[1+\frac{g^2}{2\kappa\Gamma}-
    \left(1+\frac{g^2}{\kappa\Gamma}\right)^{-1/2}
    \right] \nonumber \\
    &=& \sqrt{j_{in}\tau}\frac{g^2}{\kappa\Gamma}
     \frac{\kappa_T}{\kappa}\times
 \left\{
 \begin{array}{ll}
 1  \\
 \frac{1}{2}
 \end{array}
 \quad \mbox{ for } \left(\frac{g^2}{\kappa\Gamma}\right)
\begin{array}{ll}
 \ll 1   \\
 \gg 1
 \end{array}
 \right.
 \label{eq:Sweakav}
 \eea
This is exactly half of the maximum value of $S$ given by  Eq.\
(\ref{eq:Sweak}). The corresponding mean number $\bar{M}$ of
spontaneously scattered photons is
 \bea
 \bar{M} &=& j_{in}\tau\frac{\kappa_T}{\kappa}
    \left(\frac{\kappa\Gamma}{g^2}\right)^{1/2}
    \left(1 + \frac{\kappa\Gamma}{g^2}\right)^{-3/2}
     \\
   &=& j_{in}\tau\frac{\kappa_T}{\kappa}\times
 \left\{
 \begin{array}{ll}
 \frac{g^2}{\kappa\Gamma} \\
 \left(\frac{g^2}{\kappa\Gamma}\right)^{-1/2}
 \end{array}
  \quad \mbox{ for }\left( \frac{g^2}{\kappa\Gamma}\right)
\begin{array}{ll}
 \ll 1   \\
 \gg 1
 \end{array}
 \right.\nonumber
 \eea
We note that in the weak coupling limit, the dependence of
$\bar{M}$ on $\bar{S}$ differs from Eq.\ (\ref{eq:Mweak}) only by
a factor of two, whereas it is qualitatively different in the
strong coupling limit. The spatially averaged value $\bar{D}$ of
the momentum diffusion reads
 \be
 \bar{D} = \frac{(\hbar k)^2}{\tau}
           \left(1+\frac{g^2}{2\kappa\Gamma}\right)\bar{M}
 \ee
and thus the rms momentum change during detection is
 \be
 \Delta p = \hbar k
 \sqrt{2\bar{M}\left(1+\frac{g^2}{2\kappa\Gamma}\right)}.
 \ee
The corresponding random walk in position leads to a spatial
spreading $\Delta z$ of
 \be
 \Delta z = \sqrt{\frac{2\bar{D}}{m^2}\frac{\tau^3}{3}}
  = \frac{\Delta p}{m}\frac{\tau}{\sqrt{3}}\  ,
 \ee
where $m$ is the atomic mass. As an example, let us take the
parameters of the dotted curve in Fig.\ \ref{fig:loss} with
$j_{in}=20$ photons/$\mu$s. Then $\bar{S}=5.1$, $\bar{M}=25$,
$\Delta p=9.3\hbar k$, and $\Delta z=320$ nm. Thus the heating and
diffusive motion are relatively small and do not compromise the
feasibility of single-atom detection.

We have performed numerical simulations of a simple experiment in
which atoms move along a guide perpendicular to the cavity axis
and cross the cavity field. The parameters used are
$(g,\Gamma,\kappa_{loss}=\kappa_T)=2\pi\times(12,3,14)$ MHz,
$j_{in}=10$ photons/$\mu$s, and cavity waist $w_0=3\,\mu$m. The
transverse oscillation frequency in the guide is $2\pi\times37$
kHz, the atoms move along the guide with a mean velocity of 0.4
m/s, and the initial temperature of the cloud before interaction
with the cavity is 30 $\mu$K. Here the motion of the atoms is
classical.  It is beyond the scope of the present paper to
consider single atoms magnetically trapped in the Lamb-Dicke
regime or to analyze the effect of the cavity field on a
Bose-Einstein condensate.

\begin{figure}[tb]
 \includegraphics[height=4.5cm]{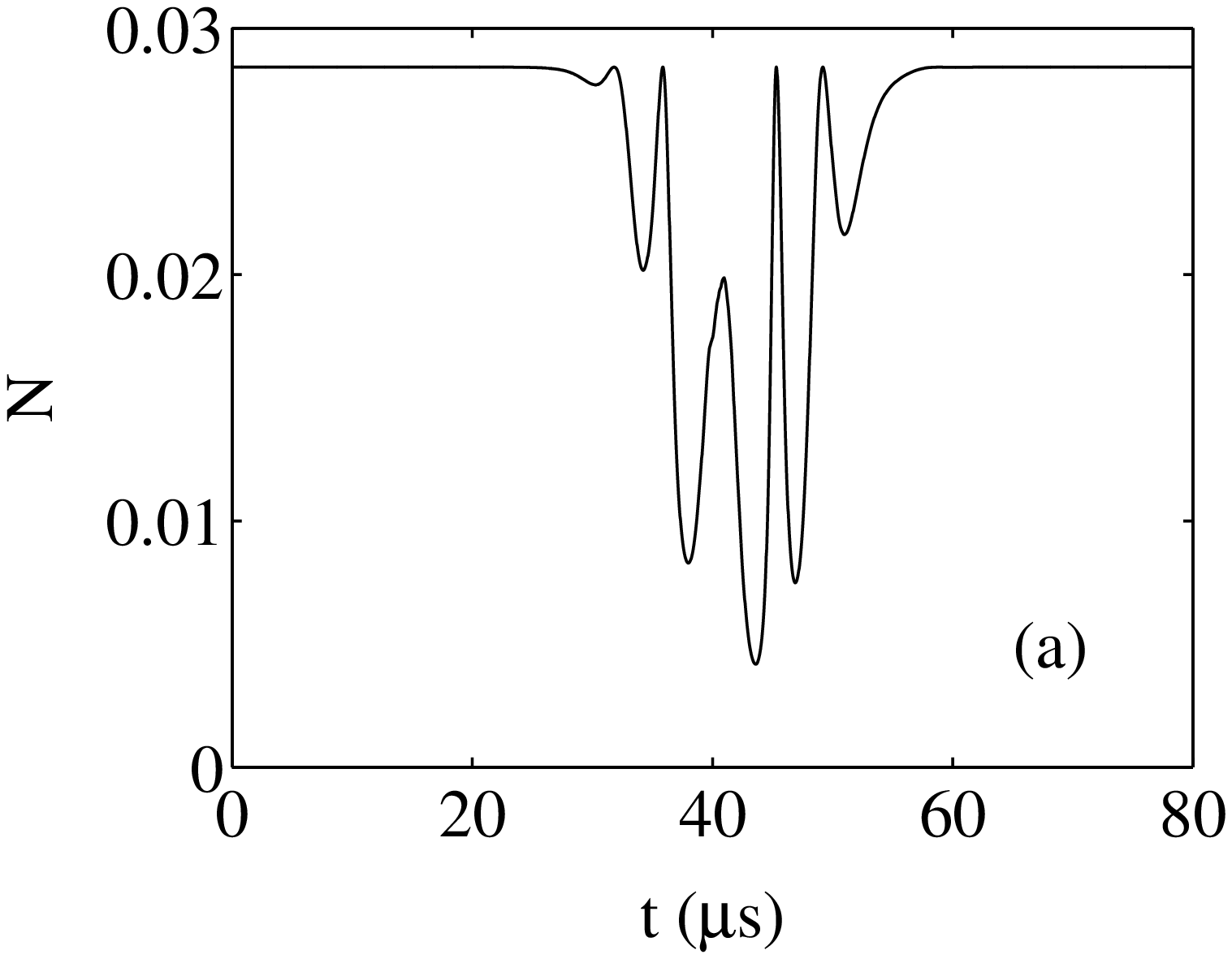}
 \includegraphics[height=4.5cm]{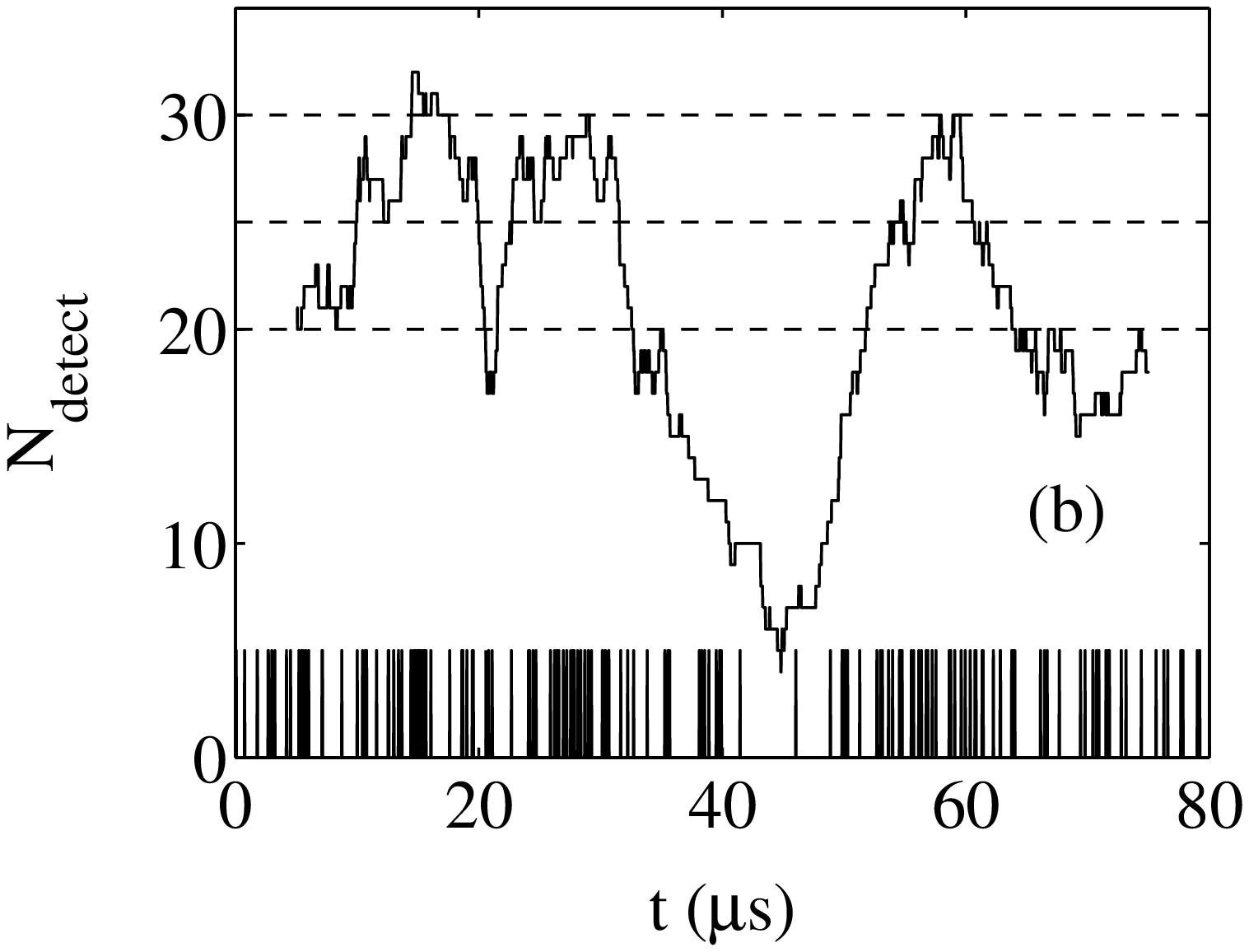}
 \caption{Numerical simulation of an atom which is trapped in a 2D harmonic trap
 but moves along the third dimension and traverses the cavity.
 (a) Stationary intra-cavity photon number $N$ vs.\ time.
 (b) Simulated detector signal vs.\ time: arrival time of individual photons at the
 photo-detector (vertical lines at bottom), number of detected photons integrated over
 intervals of 8 $\mu$s (solid line), mean integrated photon number for an empty
 cavity and corresponding quantum noise (dashed). (See text for
 details.)
 \label{fig:simu}}
\end{figure}

Figure \ref{fig:simu}(a) shows the typical temporal evolution of
the intra-cavity photon number $N$ in one particular realization
of the simulation. It is constant as long as there is no atom
interacting with the cavity mode, but with the arrival of an atom,
$N$ exhibits periods of strong reduction according to the position
of the atom within the cavity. This curve has been used to
simulate the arrival times of individual photons at the
photo-detector measuring the cavity output. Detector clicks are
indicated by the vertical lines at the bottom of Fig.\
\ref{fig:simu}(b) and the scarcity of clicks near time $t\approx
45\,\mu$s indicates the presence of the atom. This appears even
more clearly in the solid line of Fig.\ \ref{fig:simu}(b), which
plots the number of photons arriving in an $8\,\mu$s integration
window versus time. For comparison, the dashed lines show the
corresponding mean number of photons, plus and minus one standard
deviation, when there is no atom in the cavity. A simple criterion
for the detection of an atom in the cavity would thus be to define
a minimum value for $N_{detect}$ and infer the presence of an atom
whenever the curve drops below that. With the minimum set at 11,
we found that this procedure detects 77\% of the atoms, while the
shot noise of the cavity field produces only 750 false atoms per
second. Note that a higher threshold increases the detection
efficiency, but simultaneously increases the dark count rate. In
practice, a more sophisticated data analysis could provide better
discrimination between real detection events and dark counts. On
average, each atom makes $M=28.3$ spontaneous emissions, a result
quite comparable to the simple 1D averages discussed above.

%%%%%%%%%%%%%%%%%%%%%%%%%%%%%%%%%%%%%%%%%%%%%%%%%%%

\section{Micro cavity design}
\label{sec:designs}

It is beyond the scope of this paper to consider the details of
specific micro-cavity designs, but it does seem appropriate to
make some general remarks about the feasibility of building
suitable cavities.  In this work, we are not thinking of
whispering gallery cavities \cite{whisperinggalleryreference},
where the light is trapped near the perimeter of a small sphere or
disk and atoms may couple to the evanescent field that leaks out.
Rather, we have in mind an open resonator in which atoms can have
access to the regions of maximum optical field. The lower Q of the
open resonator can be compensated by the small waist size of the
cavity mode, which we are taking to be a few $\mu$m.

A simple cavity of this kind might be made using a pair of optical
fibers, each with an integrated Bragg reflector. The two fibers,
aimed at each other with a small gap between the ends, form a
cavity, see Fig.\ \ref{fig:scheme}. Of course there are
reflections at the fiber ends and losses from the region of the
gap as well as absorption in the fiber itself, but such a cavity
has the virtue that it is easy to build and its modes are readily
analyzed, which we do in appendix \ref{sec:modes}. The damping and
coupling parameters used in this work are realistic parameters for
such a cavity, as calculated in the appendix. The gap where the
atom is placed should be wide enough to avoid long-range van der
Waals or Casimir-Polder interactions between the atom and the end
of the fiber. These forces become problematic at distances below a
few hundred nm \cite{Hinds,Mabuchi}.

The cavity could be significantly improved by matching the ends of
the fibers to the wavefronts in the gap.  With this in mind we
have demonstrated a fiber terminated by a microlens that produces
a 2$\mu$m beam waist 50$\mu$m from the end of the fiber, but an
analysis of the microlens cavity is beyond the scope of appendix
\ref{sec:modes}. GRIN rod lenses also lend themselves to this
application. It is also desirable for the cavity to be tunable so
that its mode frequencies can be properly chosen relative to the
atomic transition of interest. Temperature tuning and
piezoelectric tuning are both realistic options. Alternatively,
the cavity could be tuned electro-optically using, for example, a
material such as lithium niobate (LiNbO$_3$), whose refractive
index changes by up to 1$\%$ in an applied electric field.
Ultimately it is of interest to integrate optical structures for
atom detection directly into the atom chip. Many dielectric and
semiconductor materials combine suitable optical properties with
low enough vapor pressure. Waveguides made from SiN or Ta$_2$O$_5$
could be used in the red and near infrared for atoms such as Li
and Rb, while GaAs structures would be appropriate for Cs.

%%%%%%%%%%%%%%%%%%%%%%%%%%%%%%%%%%%%%%%%%%%%%%%%%%%

\section{Conclusions}
\label{sec:conclusions}

In this work we have analyzed the use of optical microcavities for
detecting single atoms on an atom chip. We find that even cavities
of quite modest $\kappa$, in the range of $2 \pi \times 10$ MHz,
can play a useful role provided the waist of the light field is
only a few $\mu$m. With lower loss, such cavities permit the
detection of a single atom while requiring less than 1 photon to
be spontaneously scattered. The effect of optical dipole forces
can be significant.

Our calculations show that the effect on the atom is approximately
the same for resonant and far-detuned pump light if one wants to
obtain a fixed signal-to-noise ratio for the same cavity
parameters. On resonance, simple monitoring of the cavity output
power can be used to detect the atom, but very small photon
numbers must be used to achieve maximum signal-to-noise ratio. By
contrast, off-resonant detection permits much larger pump power
and therefore much larger cavity output, but in this case a more
refined homodyne technique must be used to detect the presence of
the atom in the cavity.  These methods seem promising for the
detection of single atoms on an atom chip.
%%%%%%%%%%%%%%%%%%%%%%%%%%%%%%%%%%%%%%%%%%%%%%%%%%%

%\acknowledgments

\begin{acknowledgments}

This work was supported by the European Union (Contract No.\
IST-1999-11055, ACQUIRE), the Deutsche Forschungsgemeinschaft
(Schwerpunktprogramme ``Quanteninformationsverarbeitung'',
``Wechselwirkungen in ultrakalten Atom- und Molek\"{u}lgasen''),
the Landesstiftung Baden-Wuerttemberg (Kompetenznetzwerk
``Quanteninformationsverarbeitung''), and the British Engineering
and Physical Sciences Research Council. P.\ D.\ was supported by
the European Commission (Contract No.\ HPMF-CT-2000-00788).

\end{acknowledgments}

%%%%%%%%%%%%%%%%%%%%%%%%%%%%%%%%%%%%%%%%%%%%%%%%%%%%%%%%
\appendix

\section{Modes of a fiber gap cavity}
\label{sec:modes}

\begin{figure}[tb]
 \includegraphics[height=3cm]{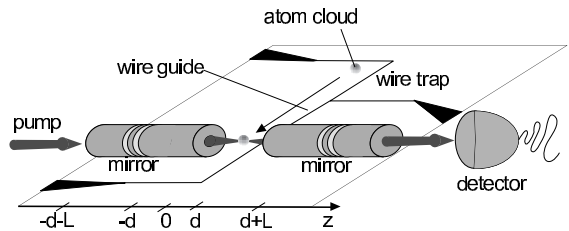}
 \caption{Schematic presentation of the fiber cavity on an atom
 chip.
 \label{fig:scheme}}
\end{figure}

In this appendix we calculate the modes and loss rates of the
simple fiber cavity used to illustrate the main part of the paper.
The cavity consists of two single-mode fibers mounted on top of
the chip and placed on opposite sides of the current carrying wire
that forms the magnetic atom trap. The guided atoms pass through a
gap of length $2d$ between the fibers, as illustrated in
Fig.~\ref{fig:scheme}. Each fiber contains a highly reflective
mirror, e.g.\ a Bragg reflector, at a distance $L$ from its end.
The cavity mode confined between these mirrors is concentrated in
the fiber cores and in the gap between the fibers. As we show
below, it is advantageous to use relatively large core diameters
to reduce the energy loss by the mode mismatch between the fibers.
The assumption of a single mode fiber thus leads to the
requirement of a very small difference of the index of refraction
between the fiber core and cladding. In this limit, the mode in
the fiber can be well approximated by a transversely polarized
field with a Gaussian profile of waist $w_0$ \cite{Agrawal}.
Similarly, the paraxial approximation can be applied to the light
field in the gap which is therefore Gaussian.

Thus, the electric field of the cavity mode can be written as
 \be
 \mathbf{E}(\mathbf{x}) = \mathbf{e_0}\times\left\{
 \begin{array}{ll}
 A_{1+}f_{1+}(\mathbf{x})+A_{1-}f_{1-}(\mathbf{x}) \quad &
 \mbox{for } z<-d,\\
 A_{2+}f_{2+}(\mathbf{x})+A_{2-}f_{2-}(\mathbf{x}) \quad &
 \mbox{for } |z|<d,\\
 A_{3+}f_{3+}(\mathbf{x})+A_{3-}f_{3-}(\mathbf{x}) \quad &
 \mbox{for } z>d
 \end{array}
 \right.
 \ee
where $\mathbf{e_0}$ is the polarization vector, $\mathbf{x}$
symbolizes the cylindrical coordinates $(r,\varphi,z)$, and
 \bea
 f_{1\pm}(\mathbf{x}) &=& e^{-r^2/w_0^2}e^{\pm ik_1(z+d)}, \\
 f_{3\pm}(\mathbf{x}) &=& e^{-r^2/w_0^2}e^{\pm ik_1(z-d)},
 \eea
are the fiber modes traveling to the right (+) and left (-) within
each of the two fibers. In the gap where the atoms are guided, the
Gaussian beams emerging from the left (right) fiber and
propagating to the right (left) are given by
 \bea
 f_{2+}(\mathbf{x}) & = &
 \frac{w_0}{w(z+d)}\exp\left[
   -\frac{r^2}{w^2(z+d)}\right. \\
   & & \left. +ik_0\frac{r^2}{2R(z+d)}+ik_0(z+d)-i\eta(z+d)
 \right], \nonumber\\
 f_{2-}(\mathbf{x}) & = & \frac{w_0}{w(z-d)}\exp\left[
   -\frac{r^2}{w^2(z-d)}\right. \\
   & & \left.-ik_0\frac{r^2}{2R(z-d)}-ik_0(z-d)+i\eta(z-d)
 \right]. \nonumber
 \eea
 Here
 \bea
 w(z) & = & w_0 \sqrt{1+(z/z_0)^2},\nonumber\\
 R(z) & = & z[1+(z_0/z)^2], \nonumber\\
 \eta(z) & = & \arctan(z/z_0), \nonumber\\
 z_0 & = & \pi w_0^2/\lambda_0.
 \eea
Because of the divergence of the Gaussian beams in the gap,
$f_{2+}$ ($f_{2-}$) does not exactly match the fiber mode $f_{3+}$
($f_{1-}$) at $z=d$ ($z=-d$). Therefore we project $f_{2\pm}$ onto
$f_{2\mp}^*$ at these positions and treat the mode mismatch as
cavity loss. We find
 \bea
 f_{2-}(r,z=-d) &=& Q f_{2+}^*(r,z=-d) + f_{2-}^{\perp}(r,z=-d), \nonumber \\
 f_{2+}(r,z=d) &=& Q f_{2-}^*(r,z=d) + f_{2+}^{\perp}(r,z=d),
 \nonumber
 \eea
where the coefficient $Q$ gives the mode-matched part of the
amplitude, while the superscript $\perp$ indicates the part of the
field that is orthogonal to the mode of the fiber and is therefore
lost.  We find that
 \be
 Q = |Q|e^{i\phi} = \frac{w_0}{w(d)}e^{ik_0 2d-i\eta(d)}.
 \ee
The phase of $Q$ is of first order in $d/z_0$, i.e., gap size
divided by twice the Rayleigh length, whereas the modulus of $Q$
is of second order. We will therefore calculate the cavity modes
in perturbation theory in $d/z_0$. The first order terms yield a
change of the optical path length and hence a change of the
resonance frequencies, but still allow for self-consistent,
lossless modes, which can then be used in the second order terms
to calculate the leading contribution of the cavity loss rate.

Using the continuity equations for the electric field at the
fiber-vacuum interface together with the boundary conditions at
the mirrors, we obtain the cavity resonance condition
 \be
 d\left(2k_0-\frac{1}{z_0}\right) =
 m\pi-2\arctan\left[\frac{1}{n}\tan(k_1 L)\right],
 \label{eq:resonance}
 \ee
where $m$ is an integer number and $n=k_1/k_0$ is the effective
refractive index of the fiber. We also obtain relations between
the electric field amplitudes $A_{j,\pm}$, in particular
 \bea
 |A_{1+}| &=& |A_{1-}| = |A_{3+}| = |A_{3-}|\\
 |A_{2+}| &=& |A_{2-}| \nonumber \\
 & =& |A_{1+}| \left| \frac{1+n-(1-n)e^{-2ik_1L}}{2}\right|.
 \label{eq:gapamp}
 \eea

The energy flow in $z$ direction in the gap is given by
 \be
 S_z = \frac{\pi w_0^2|A_{2+}|^2}{\mu_0 c}
 \ee
which leads to a loss of energy from the cavity
 \be
  S_{loss} = 2 S_z (1-|Q|^2).
 \ee
If we divide this by the total energy of the light stored in the
fiber cavity, $E=4n^2\epsilon_0\pi L w_0^2|A_{1+}|^2$ (neglecting
the small energy stored in the gap), we obtain the energy loss
rate
 \be
 2\kappa_{gap} = \frac{c}{2L}\left(\frac{d}{z_0}\right)^2 \left|
 \frac{1+n-(1-n)e^{-2ik_1L}}{2n}\right|^2
 \ee
due to the mode mismatch of the mode coupling through the gap. The
loss rate $\kappa_{loss}$ is then the sum of $\kappa_{gap}$ and
any additional loss rates, e.g., due to material absorption.
Finally, the maximum single photon Rabi frequency, observed for an
atom on the axis of the cavity in the gap, is given by
 \be
 g = \left|1+n-(1-n)e^{-2ik_1L}\right|
     \sqrt{\frac{3\Gamma c}{2n^2 L w_0^2 k_0^2}}.
 \ee

As an example let us consider rubidium atoms ($\lambda_0=780$ nm)
in such a cavity with $L=20000\lambda_1$, which yields a node of
the electric field at the fiber ends and therefore a large field
in the gap, see Eq.\ (\ref{eq:gapamp}). The gap size is chosen as
$2d=5.079\,\mu$m in accordance with the resonance condition
(\ref{eq:resonance}). The fiber core diameter is 5$\mu$m, the
refractive index of the fiber core is $n_1=1.5$, and that of the
cladding is $n_2=1.496$. For these parameters, the corresponding
waist is $w_0=2.92\,\mu$m, $\kappa_{gap}=2\pi\times 6.23$ MHz and
$g=2\pi\times 12.2$ MHz. A mirror transmission of $T=0.01$ gives
$\kappa_T=Tc/(4nL)=2\pi\times 7.65$ MHz. This example motivates
the choice of parameters used for the figures in this work. A
smaller gap of $2d=1.563\,\mu$m gives
$\kappa_{gap}=2\pi\times0.59$ MHz, which we use in the insets of
Figs.\ \ref{fig:noloss} and \ref{fig:hom}. Note that, as $L$ is
tunable, any experimental realization error concerning the exact
magnitude of $2d$ may be compensated in order to once again comply
with the resonance condition (\ref{eq:resonance}).

%%%%%%%%%%%%%%%%%%%%%%%%%%%%%%%%%%%%%%%%%%%%%%%%%%%%%%%%%%%

\end{document}